\documentstyle[12pt]{article}
\begin{document}
\vskip 4 cm
\begin{center}
\Large{\bf THE EARLY UNIVERSE AND THE STANDARD MODEL OF 
PARTICLE PHYSICS }
\end{center}
\vskip 3 cm
\begin{center}
{\bf AFSAR ABBAS} \\
Institute of Physics, Bhubaneswar-751005, India \\
(e-mail : afsar@iopb.res.in)
\end{center}
\vskip 20 mm  
\begin{centerline}
{\bf Abstract }
\end{centerline}
\vskip 3 mm

The most well tested and successful model of particle physics is the
Standard Model (SM). It is shown here that the restoration of the
full SM symmetry ( as in the Early Universe ) leads to the result that 
the electric charge loses all physical meaning and hence does not exist
above this scale, As a direct consequence of this we show that `time',
`light' along with its velocity c and therefore the theory of relativity,
all lose any physical meaning. The space-time structure as known to us,
disappears in this phase transition. Thus it is hypothesized that the
Universe came into existence when the SM symmetry $ SU(3)_c
\otimes SU(2)_L \otimes U(1)_Y $ was spontaneously broken to $ SU(3)_c
\otimes U(1)_{em} $. This does not require any spurious extensions of the
SM and in a natural and consistent manner explains the origin of the 
Universe within the framework of the SM itself. Thus the SM, in addition
to all its successes, has within itself, the capablity of giving a
consistent decription of the origin of the Universe.

\newpage

As the universe expands, it is predicted
that it undergoes a series of phase transitions [1]  during which
the appropriate symmetry breakes down in various stages until it
reaches the stage of the electro-weak (EW) symmetry.
This is given by the group structure  $SU(3)_c$
$\otimes$ $SU(2)_{L}$
$\otimes$ $U(1)_{Y} $. After $ t \sim 10^{-10} $ seconds 
( at  $T\sim 10^{2}$ GeV ) the electro-weak phase 
transition to $SU(3)_c$ $\otimes$ $U(1)_{em}$
through the Higgs Mechanism takes place. It is the strucure of this phase
transition that I look into in this paper.

Note that the structure $ SU(3)_c $ $\otimes$ $SU(2)_{L}$ $\otimes$
$U(1)_{em}$ forms the basis of the Standard Model ( SM ) of particle
physics. This model has been well studied in the laboratory. It is
the best tested model of particle physics [2]. However 
because of subjectively aesthetic reasons some people feel that
this may not be the ultimate model of physics and hence there
have been several propsals to go beyond the SM like the GUTs, 
the Superstring theories etc [2]. One has to however remember
that none of these have been shown to be correct by any empirical test in
the laboratory. Some of these models are perhaps not even testable.

The idea of the electro-weak phase transition [1,2]
is that above some critical temperature $T_{c}^{EW} $ the full electro
-weak symmetry $ SU(3)_c$ $\otimes$
$ SU(2)_{L} $$ \otimes$ $ U(1)_{Y} $ is restored. This restoration
implies that now the $ SU(2)_{L} $ gauge particles $ W^{+,-} $, 
$ W^{0} $ and the $ U(1)_{Y} $ gauge particle 
$ B_ {\mu} $ becomes massless. In addition
all matter particles $ e $, $\mu $, $ \tau $, u-quark etc becomes
massless too. All the proposed extensions presumably become operative
above this scale. As all these extensions are trying to go beyond the SM
and as the SM is the most established and tested model of particle
physics, one expects that all these extensions should be fully consistent
with the requirements of the SM. Indeed this was so in the early
70's when most of the attempts to go beyond the SM were originating [2].
It was believed then that one of the weaknesses of the SM was that the 
electric charge was not quantized in it. It was seen as a success of
the GUTs concept that the electric charge was found to be automatically
quantized therein. Hence as per popular perception the GUTs had all that
the SM had plus some more [2].

However, it was as late as 1990 that this was shown to be wrong. Against
all expectations, the electric charge was shown to be fully and
consistently quantized in the SM [3]. In fact this turns out to be its
fundamental property, no less significant than the generation of masses
for the gauge particles [3,4]. It had immediate implications for all
extensions beyond the SM [5,6,7]. If electric charge quantization in the
SM has consequences for other models, does it have more to say about the
nature of things. Here in this paper this is what shall be looked into,
and we shall find surprising and basic effects, giving us insight into
fundamental aspects of Nature. It is heartening that all this shall arise
within the SM, which as of today, has withstood closest scrutiny,
both experimental and theoretical.   

Let us start by looking at the first generation of quarks and leptons
(u, d, e,$\nu$ )  and assign them to
$SU(3)_{c} \otimes SU(2)_L \otimes U(1)_Y$ representation as follows 
[3] 

\begin{displaymath}
q_L = \pmatrix{u \cr d}_L, (3,2,Y_q)
 \end{displaymath}
\begin{displaymath} u_R; (3,1,Y_u) \end{displaymath}
\begin{displaymath} d_R; (3,1,Y_d) \end{displaymath}
\begin{displaymath} l_L =\pmatrix{\nu \cr e}; (1,2,Y_l)
\end{displaymath}
\begin{equation}
 e_R; (1,1,Y_e)
\end{equation}

To keep things as general as possible this brings in five unknown 
hypercharges.

Let us now define the electric charge in the most general way in
terms of the diagonal generators of $SU(2)_L \otimes U(1)_Y$ as
\begin{equation} Q'= a'I_3 + b'Y \end{equation}
\newline We can always scale the electric charge once as $Q={Q'\over
a'}$ and hence ($b={b'\over a'}$)
\begin{equation} Q = I_3 + bY \end{equation}

In the SM $ SU(3)_{c} $ $\otimes$ $ SU(2)_{L}$ $\otimes$
$U(1)_{Y}$ is spontaniously broken through the Higgs mechanism to the
group $ SU(3)_{c} $ $\otimes$ $U(1)_{em}$ . In this model the Higgs is
assumed to be doublet $ \phi $ with arbitrary hypercharge $ Y_{\phi}$.
The isospin $I_3 =- {1\over2}$ component of the
Higgs develops a nonzero vacuum expectation value $<\phi>_o$. Since we want
the $U(1)_{em}$ generator Q to be unbroken we require $Q<\phi>_o=0$. This
right away fixes b in (3) and we get

\begin{equation} Q = I_3 + ({1 \over 2Y_\phi})Y \end{equation}

Next one requires that the L-handed and the R-handed charges be identical
in $ U(1)_{em} $ which is a fundamental property of electrodynamics.
Also by demanding that the triangular anomaly cancels (to ensure
renormaligability of the theory) [3]
one obtaines all the unknown hypercharge in terms of the unknown Higgs
hypercharge $Y_{\phi}$. Ultimately $ Y_{\phi} $ is cancelled out
and one obtain the correct charge quantization as follows.

\begin{displaymath}
 q_L = \pmatrix{u \cr d}_L , Y_q = {{Y_\phi} \over{3}},  \end{displaymath}
\begin{displaymath} Q(u) = {2\over 3}, Q(d) = {-1\over 3}
 \end{displaymath}
\begin{displaymath} u_R, Y_u = {3\over{4}} {Y_\phi}, Q(u_R) ={2\over{3}} 
 \end{displaymath}
\begin{displaymath} d_R, Y_d = {-2\over{3}} {Y_\phi}, Q(d_R) ={-1\over3} 
 \end{displaymath}
\begin{displaymath} l_L = \pmatrix{\nu \cr e}, Y_l = -Y_\phi,
Q(\nu) = 0, Q(e) = -1
  \end{displaymath}
\begin{equation}
 e_R, Y_e = -2Y_\phi, Q(e_R) = -1
\end{equation}

It has been shown [3,4] that  for arbitrary $ N_{c} $ the colour
dependence of the electric charge as demanded by the SM is 

\begin{displaymath}
\newline Q(u) = {1\over 2}(1+{1\over N_c})
\end{displaymath}
\begin{equation}
\newline Q(d) = {1\over 2}(-1+{1\over N_c})   
\end{equation}

One should note that equations  (5) and (6)
show that  the electric charge is
quantized in SM. The complete structure of the SM as is, is required
to obtain this result on very general grounds. The SM is the best
tested model of particle physics. What has this to
say about the early universe scenarios available today ?

Clearly the $ U(1)_{em}$ symmetry which
arose due to spontaneous symmetry breaking due to a Higgs doublet
in the EW symmetry will be lost above $ T_{c}^{EW} $ whence
$ SU(2)_{L}$ $\otimes$ $ U(1)_{em}$ symmetry would be
restored. As is obvious, above $ T_{c}^{EW}$ all the fermions and gauge bosons
becomes massless . This properly is well-known and
has been incorporated in cosmological models. Here I point out a new 
phenomenon arising from the restoring of the full EW symmetry .

Note that to start with the parameter b and Y in equation (3) in the
definition
of electric charge were completely unknown. We could lay a handle on 'b'
entirely on the basis of the presence of spontaneous symmetry breaking
and on ensuring that photon was massless $ b = \frac{1}{2 Y_{\phi}} $.
Above $ T_{c}^{EW} $ where the EW symmetry is restored there is
no spontaneous symmetry breaking and hence the parameter b is
completely undetermined. Together 'bY' could be any arbitrary number
whatsoever even an irrational number. Within
the framework of this model above $ T_{c}^{EW} $ we just cannot define electric
charges at all. It may be a number which is zero or infinite
or an irrational etc. Hence the electric charge given by equation (3)
loses any physical meaning all together.This is the new amazing result.

So above $ T_{c}^{EW} $ all the particles have not only become massless,
they have forgotten their charges also. It just does not make sense to talk
of their charges. The concept of electric charge has been lost.
There is no such thing as charge anymore.
The photon (which was a linear combination of $W^{0}$
and $ B_{\mu} $ after spontaneous symmetry breaking ) with it's
defining  vector
characterteristic  does not exist either. So the conclusion is that 
there is no electrodynamics above $ T_{c}^{EW} $.

Note that the electric charge in the SM was not an elementary or 
fundamental object at all. In fact it was
a secondary quantity defined in terms of the elementary objects
$ I_{3}$ and Y ( see equation(3)). So it should not be really 
surprising to see it loose it's meaning under special circumstances.
The same is true
of the photon of $ U(1)_{em} $ . In short $ U(1)_{em} $ owes its
existence to SSB and looses it's meaning when 
the full EW symmetry is restored [8,9]. 

I have demostrated that actually electric charge is quantised in
SM and when at high temperature the EW symmetry is
unbroken, the concept of electric charge does not arise. At those and
still higher temperatures ie at still earlier universe 
[1] extensions like GUTs, Supergravity, 
Superstrings are believed to be relevant. 
Note that in all these models, as a generic property 
charge is always believed to exist and is quantized [8,9].
Quite clearly the existence of electric charge demanded by these 
models in the very early universe is in  conflict with what I have
shown that there was no elctric charge before about $ 10^{-10} $
seconds. Hence all these extensins can not give a valid description
of the very early universe [8,9].

All this misunderstanding and confusion arose because we had not understood
the electro-weak phase transition correctly. We have already seen that the
restoration of the full SM symmetry leads to the result that there
is no electric charge above it and also that there was no photon.
All the so called hypothetical extenions beyond the SM are already ruled
out as shown above. So what remains?

The above result shows that above the electro-weak  phase transition there
was no electromagnetism. Maxwells equations of electrodynamics show that
light is an electromagnetic phenomenon. Hence above the electro-weak phase
transition there was no light and no Maxwells equations. And surprisingly
we are led to conclude that there was no velocity of light c as well. One
knows that the velocity of light c is given as
 
\begin{equation}
c^2 = { 1 \over { \epsilon_0 \mu_0} }
\end{equation}

where $ \epsilon_0  $ and $ \mu_0 $ are permittivity and permeability
of the free space. These electromagnetic properties disappear
along with the electric charge and hence the velocity of light also
disappears above the electro-weak phase transition.

The premise on which the theory of relativity is based is 
that c, the velocity of light is always the same, no matter from which
frame of reference it is measured. Relativity theory also asserts that
there is no absolute standard of motion in the Universe. Of
fundamental significance is the invariant interval

\begin{equation}
s^2 = { ( c t ) }^2 - x^2 - y^2 - z^2
\end{equation}

Here the constant c provides us with a means of defining time in terms of
spatial separation and vice versa through l = c t. this enables one to
visualize time as the fourth dimension. Hence time gets defined due to
the constant c. Therefore when there were no c as in the early universe 
,there was no time as well. Hence above the electro-weak breaking scale 
there was no time. As the special theory of relativity depends upon c and
time, above the electro-weak breaking scale
there was no special theory of relativity. As the General Theory of
relativity also requires the concept of a physical space-time, it
collapses too above the electro-weak breaking scale.
Hence the whole physical universe collapses above this scale. 

Therefore it is hypothesized here that the Universe came into existence
when the electro-weak symmetry was broken spontaneously.
Before it there was no 'time', no 'light', no maximum and constant
velocity like 'c', no gravity or space-time structure on which objective
physical laws as we know it could exist.

Note that this new picture of the origin of the Universe arises naturally
within the best studied and the best verified 
model of particle physics. It does not require any
ad-hoc or arbitrary models or principles. One only has to do a consistent 
and careful analysis of the hitherto misunderstood electro-weak
spontaneous symmetry breaking.

In short we find that the SM of particle physics is not only good to
explain all known facts, it is also capable of giving a natural and
consistent description of the origin of the Universe. It is beautiful as
it is so simple. This work should be seen as further strengthening the
SM.
 
\newpage

\vskip 4 cm
\begin{center}
{\bf\large REFERENCES }
\end{center}

\vskip 2 cm
1. E. W. Kolb and M. S. Turner, " The Early Universe "
Edison Wesley, New York ( 1990 )

2. T. P. Cheng and L. F. Lee, " Gauge Theory of Elementary Particle
Physics ", Clarendon Press, Oxford ( 1988 )

3. A. Abbas, 
{\it Phys. Lett. }, {\bf B 238} ( 1990) 344

4. A. Abbas, 
{\it J. Phys. G. }, {\bf 16 } ( 1990 ) L163

5. A. Abbas, {\it Nuovo Cimento }, {\bf A 106} ( 1993 ) 985

6. A. Abbas, {\it Hadronic J.}, {\bf 15} ( 1992 ) 475

7. A. Abbas, {\it Ind. J. Phys.}, {\bf A 67} ( 1993 ) 541

8. A. Abbas,
{\it Physics Today } ( July 1999 )  p.81-82

9. A. Abbas,
{\it `On standard model Higgs and superstring theories'} 
"Particles, Strings and Cosmology PASCOS99", Ed K Cheung,
J F Gunion and S Mrenna, World Scientific, Singapore ( 2000) 123

\end{document}